\documentclass[fleqn,twoside]{article}
\usepackage{espcrc2}

\usepackage{graphicx,epsfig}
\usepackage[figuresright]{rotating}

\newcommand{\AmS}{{\protect\the\textfont2
  A\kern-.1667em\lower.5ex\hbox{M}\kern-.125emS}}

\title{Exotic hybrid mesons from improved Kogut-Susskind fermions}

\author{
    C. Bernard \address{Department of Physics, Washington University,
St.~Louis, MO 63130, USA},
    T. Burch \address[ARIZ]{Department of Physics, University of Arizona,
Tucson, AZ 85721, USA},
    C. DeTar \address[UTAH]{Physics Department, University of Utah, Salt Lake
City, UT 84112, USA},
    Steven Gottlieb \address{Department of Physics, Indiana University,
Bloomington, IN 47405, USA},
    E.B. Gregory \addressmark[ARIZ]\thanks{Presented by E.B. Gregory},
    U.M. Heller \address{CSIT, Florida State University, Tallahassee, FL
32306-4120, USA},
    J. Osborn \addressmark[UTAH],
    R. Sugar \address{Department of Physics, University of California, Santa
Barbara, CA 93106, USA},
    and
    D. Toussaint \addressmark[ARIZ]
}
       
\begin{document}

\begin{abstract}
We summarize our measurement of the mass of the exotic $1^{-+}$ hybrid 
meson using an improved Kogut-Susskind action. We show results from both 
quenched and dynamical quark simulations and compare with results from 
Wilson quarks. Extrapolation of these results to the physical quark mass
allows comparison with experimental candidates for the $1^{-+}$ hybrid meson.
\vspace{1pc}
\end{abstract}

\maketitle
\section{INTRODUCTION}
Experimental evidence suggests the existence of particles with ``exotic'' 
quantum numbers $J^{PC}=1^{-+}$, such as the $\pi_1(1400)$ \cite{pi1400} 
and the $\pi_1(1600)$ \cite{pi1600}. Explanations of these include four-quark 
states and hybrid mesons --- mesons with gluonic excitations.
Several lattice studies \cite{Lacock:1996ny,Bernard:1997ib,ZSU} 
have used quenched Wilson and improved Wilson fermions 
to explore the possibility of such hybrid states. Using both quenched and 
dynamical Kogut-Susskind quarks, the MILC Collaboration has studied the 
$1^{-+}$ hybrid closer to the physical quark mass limit.

\section{$1^{-+}$ HYBRID MESON OPERATOR}
We can construct a $1^{-+}$ hybrid meson operator as the cross product of 
a color octet rho meson and the chromomagnetic field: 
$\rho \times B$ \cite{Bernard:1997ib}. We have several 
choices of rho meson operators, but it is convenient to 
choose the K-S flavor singlet  
$\rho_s$, with the spin $\otimes$ flavor structure
$\gamma_i \otimes {\mathbf 1}$. 
This is because each spin component of the $1^{-+}$ includes two terms, for 
example:
\begin{equation}
1^{-+}_x = \rho_y B_z - \rho_z B_y.
\end{equation}
If the flavor of the $\rho$ is dependent on its spin state, the resulting 
contribution is not a flavor eigenstate.





\begin{figure}[t]
\begin{center}\scalebox{0.3}{
\includegraphics{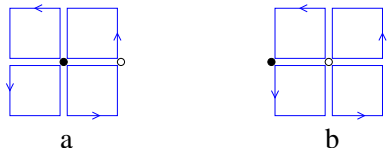}}
\end{center}
\vspace*{-0.5in}
\caption{\label{fields} Chromomagnetic field measured at the quark (a) 
and at the antiquark (b).}
\vspace{-0.2in}
\end{figure}

In computing the field strength, we apply 32 APE smearing \cite{APE}
iterations to the spatial gauge field links.
There is a subtlety in using the chromomagnetic field to generate the 
full $1^{-+}$ hybrid source operator. We might first determine the 
magnetic field strength everywhere, then multiply at each site by the 
quark source 
vector, and finally symmetrically shift to get the appropriate antiquark 
source vector.
Alternately, we might first apply the symmetric shift to the quark source 
to form an antiquark source, then multiply by the value of the 
chromomagnetic field.
These methods are equivalent to measuring the $B$-field at 
the site of the quark (Fig. \ref{fields}a) and antiquark 
(Fig. \ref{fields}b), 
respectively. Because the quark and 
antiquark are spatially separated in the flavor singlet $\rho_s$, neither of 
these represents an eigenstate of charge conjugation. To get an eigenstate of 
$C$, we use a symmetrized combination of these for the $1^{-+}$ hybrid 
operator.

\section{SIMULATION \& MEASUREMENT}
We measured the connected correlator of the $1^{-+}$ hybrid state on three 
sets of $ 28^3 \times 96 $ lattices generated with 
the ``$a^2_{\rm tad}$'' action \cite{Asqtad}. To isolate the effects of 
dynamical quarks, we used matched quenched and full QCD lattices with
$\beta=8.40$, $m_{\rm val}a=0.016,0.04$, for the quenched quarks, $\beta=7.18$ for lattices with 
three degenerate flavors of dynamical sea quarks at the strange quark 
mass ($ma=0.031$) and  $\beta=7.11$ for lattices with 
$m_{u,d}=0.4m_s$ ($ma=0.0124$). These
choices of $\beta$ give approximately the same lattice spacing 
($\sim 0.09$ fm) in the three cases.  The corresponding choices of quark mass
allow simulation at roughly equivalent values of $(m_{\rm PS}/m_V)^2$, 
the square of the ratio of the psuedoscalar to vector meson masses.

\section{RESULTS}
We fit the measured correlators to the sum of oscillating and normal 
exponentials:
\begin{eqnarray}
C(t)=A_1e^{-M_{1^{-+}}t}&+& A_2(-1)^te^{-m_2t} \nonumber\\
&+& A_3(-1)^te^{-m_3t},
\end{eqnarray}
where $M_{1^{-+}}$ is the hybrid meson mass of interest and $m_2$ and $m_3$ are
masses of nonexotic parity partner states which have oscillating correlators in
the Kogut-Susskind formulation.
We performed both four and five parameter fits. For the four parameter fits,
we fix $A_3=m_3=0$.  For the five parameter fits we fix $m_3$ to a 
pre-determined $a_1$ meson mass and fit for $A_3$ as well. We varied the 
range of the fit and tried to choose values for $M_{1^{-+}}$ corresponding to 
high-confidence fits that were insensitive to $t_{\rm max}$ 
and $t_{\rm min}$, the limits of the fit range. 
\begin{figure}[t]
\rule{0.0in}{0.3in}\vspace*{-0.0in}\\
\epsfxsize=3.10in
\epsfbox[0 0 4096 4096]{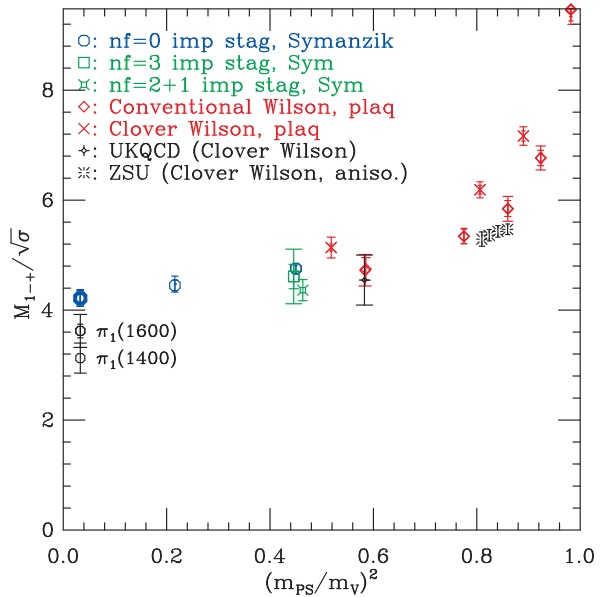} 
\rule{0.0in}{0.01in}\vspace{-1.0in}\\
\caption{Summary of $1^{-+}$ hybrid meson mass predictions as a function
of $(m_{\rm PS}/m_V)^2$. The bold octagon represents the linear 
extrapolation of $n_f=0$ data to $(m_{\rm PS}/m_V)^2=0.033$.
\label{all_results}}
\vspace{-0.25in}
\end{figure}

In Fig. \ref{all_results} we summarize our results  along with the results 
of previous Wilson quark studies by the MILC group \cite{Bernard:1997ib} 
and the UKQCD collaboration
\cite{Lacock:1996ny} as well as recent results from the Zhongshan University 
group \cite{ZSU} using Wilson quarks on an anisotropic lattice.
We use the string tension $\sigma$ to establish the lattice length scale 
and plot $M_{1^{-+}}/\sqrt{\sigma}$. For comparison, we include the 
$1^{-+}$ experimental candidates $\pi_1(1400)$ and $\pi_1(1600)$ 
at the physical value of $(m_{\rm PS}/m_V)^2=(m_{\pi}/m_\rho)^2=0.033$. We
use $\sqrt{\sigma}=440\pm 38 {\rm MeV}$ \cite{Teper:1997am} for the 
vertical scale.

\begin{figure}[t]
\vspace{-1.0in}
\epsfxsize=2.8in
\epsfbox[0 0 4096 4096]{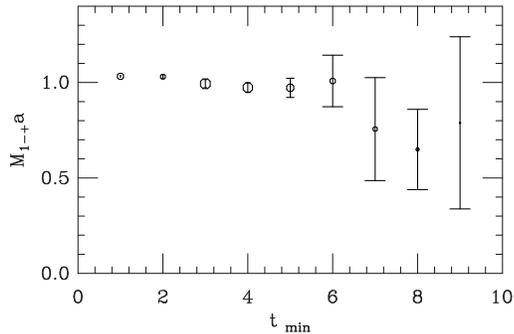} 
\vspace{-0.5in}
\caption{$M_{1^{-+}}$ {\em vs.} $t_{\rm min}$ for $\beta=8.40$ quenched lattices, 
$t_{\rm max}=15$, 5-parameter fit. Symbol size is proportional to confidence 
level.
\label{stability}}
\vspace{-0.3in}
\end{figure}

For the quenched lattices we were able to fit the data with reasonable 
confidence levels (25-50\%) for valence quark masses $ma=0.016$ and $ma=0.040$.
Fig. \ref{stability}
shows an example of fits for the quenched lattices for $t_{\rm max}=15$.
We performed a linear extrapolation of these results to the physical
value of  $(m_{\rm PS}/m_V)^2$. (Fig. \ref{all_results}.)

For lattices with three degenerate sea quarks at $m_s$, we were also able to 
extract a value
for $M_{1^{-+}}$ in reasonable agreement with the quenched result. The fits, 
however, exhibited slightly larger statistical errors, and a slight 
dependence on range. 

The lattices with $m_{u,d}=0.4m_s$ proved more interesting. 
In the case of a valence quark mass equal to $m_s$ 
($ma=0.031$), we use the masses of the $s\bar{s}$ pseudoscalar state 
and the vector $\phi$ state, measured on the same lattices, for 
$m_{\rm PS}$ and $m_V$ respectively. The fitted mass agrees with those of the 
quenched and three-flavor results within two standard deviations, but with 
larger systematic errors, estimated from the dependence on fit range.

In the case of the light valence quark 
($ma=0.0124$), we were unable to say much about the $1^{-+}$ hybrid mass
with any confidence.  The fits were very range dependent, indicating 
the likely presence of four-quark states into which the hybrid can decay.

\section{DISCUSSION AND CONCLUSIONS}
Interpretation of lattice results depends crucially on understanding the 
lattice length scale. We use the string tension 
$\sigma$ as opposed to other static potential scales such as $r_0$ or $r_1$
for two reasons. First, hybrid mesons are extended objects that feel the 
linear part of the static quark potential, where $\sigma$ is defined, 
more than the Coulombic part, where $r_0$ is defined. Second, using $\sigma$
to define the lattice spacing brings quenched and dynamical quark
data into closer agreement for both the hybrids and for low-lying non-exotic 
hadrons \cite{MILC_spectrum}. Note that if we use $r_1=0.34$ fm, our quenched 
data extrapolates to 2033(70) MeV, whereas using $\sqrt{\sigma}=440$ MeV we get
1854(65) MeV, quoting statistical errors only.

The $m_{u,d}=0.4m_s$ data illustrates that dynamical quarks introduce 
new and significant processes that contribute to the $1^{-+}$ propagator. 
Mixing with four-quark states is one possibility.  We now
have ahead of us the task of understanding these contributions so that we 
can make useful predictions of the $1^{-+}$ hybrid mass in the presence of 
dynamical quarks.

\section{ACKNOWLEDGEMENTS}
Computations for this work were performed at SDSC, PSC, ORNL and NERSC. 
This work was supported by the U.~S. ~DOE and NFS.

\end{document}